\documentclass[a4paper]{jpconf}
\usepackage{amssymb}
\usepackage{graphicx}

\begin{document}
\title{Ab initio calculation of the Hoyle state and a new look at clustering
  in nuclei}

\author{Ulf-G. Mei\ss ner}

\address{Universit\"at Bonn, HISKP and Bethe Center for Theoretical Physics,
  D-53115 Bonn, Germany\\ 
  Forschungszentrum J\"ulich, IKP-3, IAS-4 and JCHP, D-52425 J\"ulich, Germany}

\ead{meissner@hiskp.uni-bonn.de}

\begin{abstract}
I present an {\sl ab initio} calculation of the spectrum of $^{12}$C,
including also the famous Hoyle state. Its structure is discussed and
a new interpretation of clustering in nuclear physics is given.
\end{abstract}

\section{What is so special about the Hoyle state?}

In 1953, Fred Hoyle analyzed the production of heavy elements like $^{12}$C
or $^{16}$O in massive stars. He found that by far too few $^{12}$C was
formed if the so-called triple-alpha process would proceed via the ground
state of the carbon nucleus. Consequently, Hoyle predicted a $0^+$-resonance
in the vicinity of the $^4$He plus $^8$Be threshold that increases the
carbon production by many orders of magnitude~\cite{Hoyle:1954zz}. This
excited state was firmly established experimentally at Caltech in 1957
\cite{Cook:1957zz}. Ever since, the Hoyle state has been an enigma for
nuclear structure theory. For example, ab initio calculations\footnote{An 
{\sl ab initio} calculation is defined as follows: 
First, one determines the parameters that appear in the nuclear force 
{\sl entirely} from a fit to  observables in systems with $A\leq 4$. Then, the 
so-determined forces are used without modification in an exact solution of 
the $A$-body problem, $A\geq 5$.} based on 
the successful no-core-shell-model utilizing modern nuclear forces 
between two and three nucleons do very well in 
the description of the spectra of p-shell nuclei but fail miserably with
respect to the Hoyle state (or its analogs) , see~\cite{Navratil:2007we,Roth:2011ar}.
On the contrary, alpha cluster models as discussed in this conference
by Martin Freer \cite{Freer} are quite successful in describing the 
Hoyle state and other similar excited states in $^{12}$C and $^{16}$O,
but with little connection to the underlying few-nucleon forces (see
also the recent review \cite{Yamada:2011bi}). In such type of models, 
the Hoyle state has a spatial extension that is more than 1.5 times
larger than the ground state which seems to be consistent with precision
data on electromagnetic transitions, see e.g. Ref.~\cite{Chernykh:2007zz}.

In this talk, I will report on the first ever ab initio calculation of the
Hoyle state and discuss a new look on the phenomenon of clustering in nuclei,
which emerges naturally in the framework of nuclear lattice simulations.
This new approach to the nuclear many-body problem is based on the 
forces derived from the chiral effective Lagrangian of Quantum Chromodynamics
(QCD).  For few-nucleon systems, the chiral effective field theory
(EFT) for the forces between two, three and four nucleons has been worked out up to 
next-to-next-to-next-to-leading order (N$^3$LO) in the chiral power
counting for the nuclear potential. This potential  consists of 
long-range pion exchange(s) and shorter ranged multi-nucleon contact interactions.
Bound and scattering states are calculated based on exact solutions of
the Lippmann-Schwinger or Faddeev-Yakubowsky equations.
This method has passed many tests as reviewed in Ref.~\cite{Epelbaum:2008ga}. 
To tackle systems with more than four nucleons, a novel scheme that combines 
these forces with Monte Carlo methods that are so successfully used in lattice 
QCD was recently developed. This novel scheme is
termed ``nuclear lattice simulations''. For its foundations and early
applications, see the review by Lee~\cite{Lee:2008fa}.

%%%%%%%%%%%%%%%%%%%%%%%%%%%%%%%%%%%%%%%%%%%%%%%%%%%%%%%%%%%%%%%%%%%%%%%%%%
\section{The framework of nuclear lattice simulations}

\begin{figure}[b]
\begin{minipage}{36pc}
\vspace{4mm}
\includegraphics[width=0.48\textwidth]{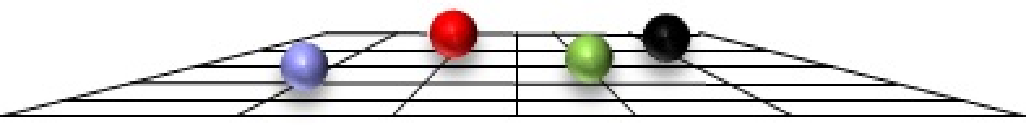}\hspace{1pc}
\includegraphics[width=0.48\textwidth]{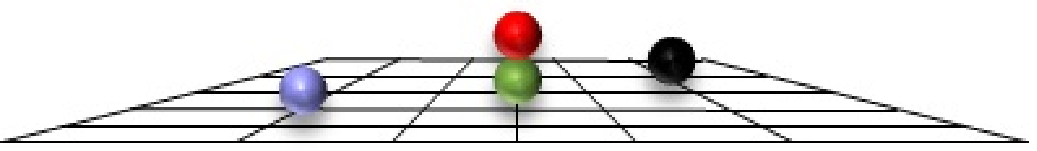}\\
\includegraphics[width=0.48\textwidth]{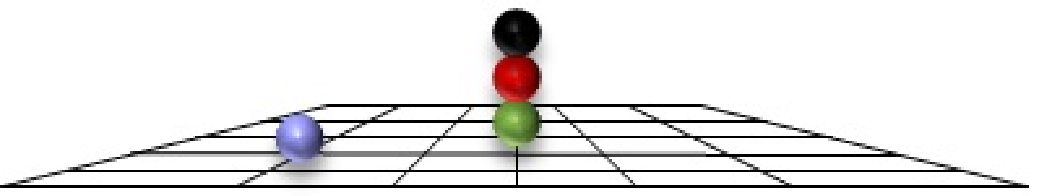}\hspace{1pc}
\includegraphics[width=0.48\textwidth]{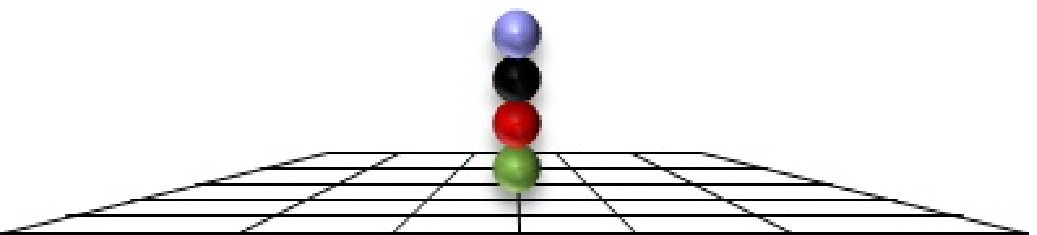}

\vspace{-2mm}
\caption{\label{fig:configs} Topologically different configurations
to put nucleons on a lattice. From the left upper to the lower right 
corner: all four nucleons on different sites, two/three/four nucleons
on one site, respectively.}
\end{minipage}\hspace{2pc}%
\end{figure}

Here, I very briefly review the framework of nuclear lattice
simulations. For more details, I refer e.g. to 
Refs.~\cite{Lee:2004si,Borasoy:2006qn,Borasoy:2007vi,Epelbaum:2009zsa}.
To be able to evaluate the generating functional 
$Z_A = \langle \psi_A | \exp(-t H)|\psi_A\rangle$ by numerical
methods, space-time is discretized in Euclidean time $t$ on a torus of volume
$L_s\times L_s\times L_s\times L_t$, with $L_s (L_t)$ the side 
length in spatial (temporal) direction. $H$ is the nuclear Hamiltonian
derived from the chiral effective Lagrangian of QCD and $\psi_A$ is a
Slater determinant for $Z$ protons and $N$ neutrons, with $Z+N=A$.
The minimal distance
on the lattice, the so-called lattice spacing, is $a$ ($a_t$)
in space (time). This entails a maximum momentum on the lattice,
$p_{\rm max} = \pi/a$, which serves as an UV regulator of the theory.
A typical lattice spacing used in the actual simulations is $a \simeq 2\,$fm,
corresponding to a momentum cut-off of about 300~MeV. In contrast to
lattice QCD, the continuum limit $a \to 0$ is not taken.
The nucleons are point-like particles residing on the lattice sites,
where as the nuclear interactions (pion exchanges and contact terms)
are represented as insertions on the nucleon world lines using standard
auxiliary field representations. For the further discussions, it 
is important to realize which configurations to put nucleons on
the lattice are possible. These are displayed in Fig.~\ref{fig:configs}.
Of particular importance are the configurations with four fermions on
one site in harmony with the Pauli principle. In fact, in a leading order
calculation using the two independent four-nucleon contatct operators without 
derivatives $\sim (\psi^\dagger \psi)^2$, one finds that in the $^4$He system the
ground state is severely overbound and consists almost entirely of the quantum
state with all four nucleons occupying the same lattice site. This is in part
due to a combinatorial enhancement of the contact interactions when more than two
nucleons occupy the same lattice site. This effect is partly overcome by higher
order four-nucleon operators, but it is most efficiently dealt with by
a Gaussian smearing procedure, which turns the point-like vertex into an
extended structure. For more details, see Ref.~\cite{Borasoy:2006qn}.
There is one further issue to be discussed. The nuclear forces have an approximate
spin-isospin SU(4) symmetry (Wigner symmetry) \cite{Wigner:1936dx}
that is of fundamental importance in suppressing the malicious sign oscillations 
that plague any Monte Carlo (MC) simulation of strongly interacting fermion systems 
at finite density. 
For this reason, nuclear lattice simulations allow access to a large part
of the phase diagram of QCD, where as 
calculations using  lattice QCD are limited to finite temperatures and
small densities (baryon chemical potential).

%\begin{figure}[tb]
%\begin{minipage}{36pc}
%%\vspace{2mm}
%\includegraphics[width=0.46\textwidth]{pipispec.eps}\hspace{2pc}
%\includegraphics[width=0.52\textwidth]{pipimsp.eps}
%\caption{\label{fig:Yeta}Left: Prediction for the $\pi^+\pi^-$ mass
%  distribution in the molecular picture of the $Y(4660)$ (solid line: central
%  parameters, shaded area: theoretical uncertainty) compared to the data 
%  \cite{Yexp}. Right: Prediction for the $\pi^+\pi^-$ mass
%  distribution in the molecular picture of the $Y_\eta$.}
%\end{minipage}
%\end{figure}

%%%%%%%%%%%%%%%%%%%%%%%%%%%%%%%%%%%%%%%%%%%%%%%%%%%%%%%%%%%%%%%%%%%%%%%%%%
\section{Ab initio calculation of the carbon-12 spectrum}

So far, we have performed calculations at N$^2$LO in the chiral expansion
of the nuclear potential. In the two-nucleon system, we have nine parameters
that are determined from a fit to the S- and P-waves in $np$ scattering. Two further
isospin-breaking parameters are determined from the $pp$ and $nn$ scattering
lengths. The Coulomb force between protons is also included.
The three-nucleon force features only two low-energy constants,
that can be determined from the triton binding energy and low-energy 
neutron-deuteron scattering in the doublet channel \cite{Epelbaum:2009zsa}.
The first non-trivial predictions of our approach are the energy dependence of the $pp$
$^1S_0$ partial wave, which agrees with the Nijmegen partial wave analysis up to momenta
of about the pion mass and the binding energy difference between the
triton ($^3$H ) and $^3$He. We find $E(^3{\rm H}) - E(^3{\rm He}) =
0.78(5)~{\rm MeV}$, in good agreement with the experimental value of
$0.76$~MeV \cite{Epelbaum:2009pd,Epelbaum:2010xt}. The ground state energies of nuclei
with $A = 4,6,12$ where also calculated in these papers.

Our NLEFT collaboration\footnote{NLEFT stands for Nuclear
  Lattice Effective Field Theory.} has recently completed {\sl ab initio} lattice calculations
of the low-energy spectrum of $^{12}$C using chiral nuclear EFT
\cite{Epelbaum:2011md}. In addition to the ground state and the
excited spin-2 state\footnote{The orientation bias in our initial state
ensemble leads to the splitting of the $J_z=0$ and $J_z=2$ components of the
spin-2 excited state.}, 
we find a resonance with angular momentum zero and positive parity at
$-85(3)$~MeV, very close to the $^4$He+$^8$Be threshold at $-86(2)$~MeV.
This is nothing but the Hoyle state. To arrive at these results, it was
of prime importance to further improve the action as compared to the
earlier ground state energy calculations. In particular,  the
effects of the finite lattice spacing and the unphysical mixing of partial waves
due to the breaking of the rotational symmetry was minimized. In addition, we have developed
a multi-channel projection Monte Carlo technique that allows to extract
excited states from combinations of initial standing waves with zero total
momentum and even parity.
\renewcommand{\arraystretch}{1.2}
\begin{table}[t!]
\begin{center}
\begin{tabular}
[c]{|l||c|c|c|c|}\hline
& $0_{1}^{+}$ &$0_{2}^{+}$ & $2_{1}^{+}$, $J_{z}=0$ & $2_{1}^{+}$, $J_{z}=2$\\\hline\hline
LO [$O(Q^{0})$] & $-110(2)$ & $-94(2)$ & $-92(2)$ & $-89(2)$\\\hline
NLO [$O(Q^{2})$] & $-93(3)$ & $-82(3)$ & $-87(3)$ & $-85(3)$\\\hline
IB + EM [$O(Q^{2})$] & $-85(3)$ & $-74(3)$ & $-80(3)$ & $-78(3)$\\\hline
NNLO [$O(Q^{3})$] & $-91(3)$ & $-85(3)$ & $-88(3)$ & $-90(4)$\\\hline
Experiment & $-92.16$ & $-84.51$ & \multicolumn{2}{|c|}{$-87.72$}\\\hline
\end{tabular}
\end{center}
\caption{Lattice results for the ground state $0_{1}^{+}$
and the low-lying excited  states of $^{12}$C. For comparison 
the experimentally observed energies are shown. All energies are in units of MeV.}
\label{excited states}%
\end{table}
In Table~\ref{excited states}, I show results for ground state and
the low-lying excited states
of $^{12}$C at leading order (LO), next-to-leading order (NLO),
next-to-leading order with isospin-breaking and electromagnetic corrections
(IB + EM), and next-to-next-to-leading order (NNLO).  
%All energies are in units of MeV. 
For comparison, the experimentally observed energies are also
listed. The error bars in Table~\ref{excited states} are one standard
deviation estimates which include both Monte Carlo statistical errors and
uncertainties due to extrapolation at large Euclidean time.  Systematic
errors due to the omitted higher-order interactions can be estimated from the
size of corrections from $O(Q^{0})$ to $O(Q^{2})$ and from $O(Q^{2})$ to
$O(Q^{3})$.  
\begin{figure}[t]
\begin{center}
\includegraphics[width=0.50\textwidth]{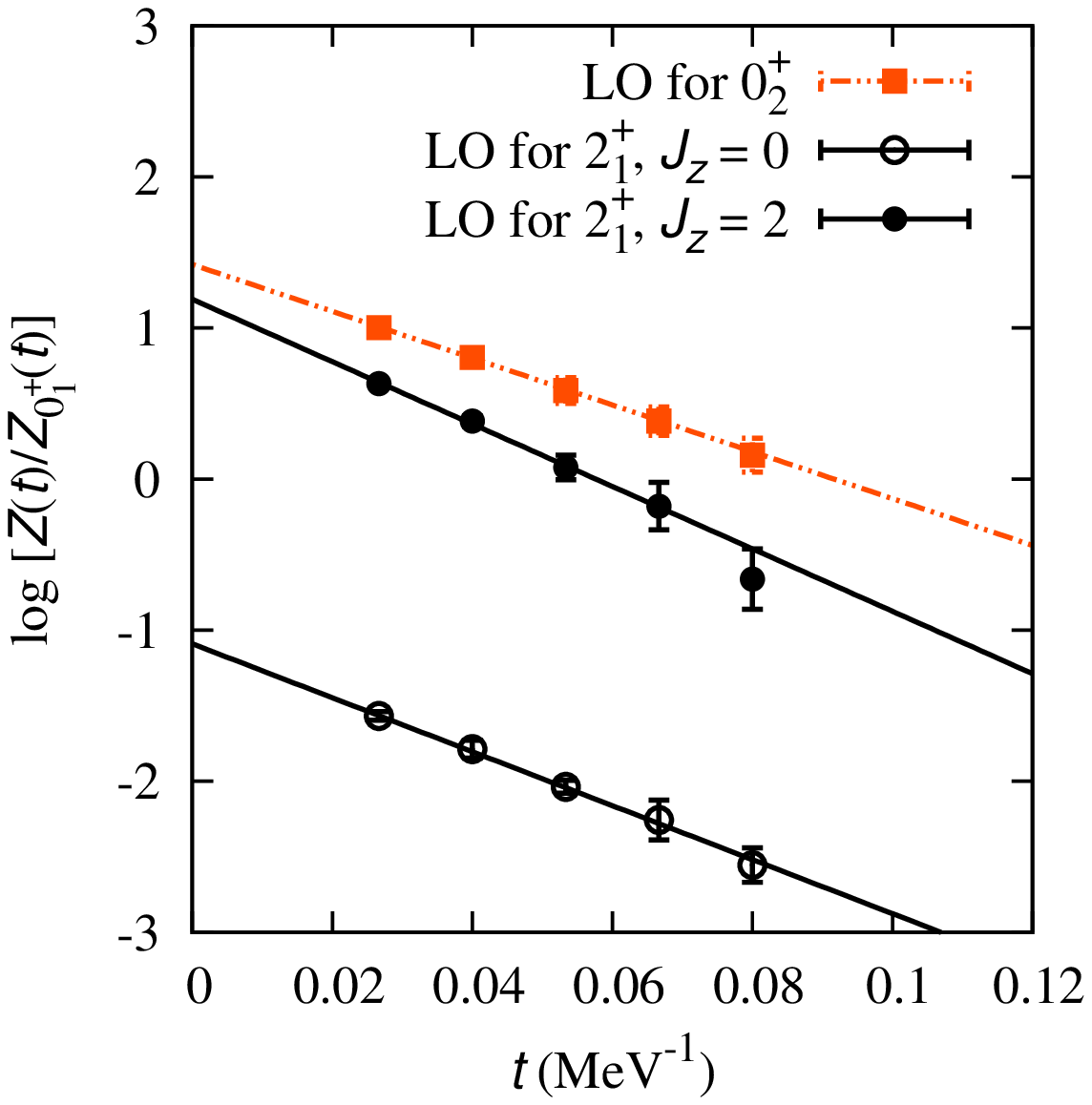}
\hspace{0.07cm}\includegraphics[width=0.48\textwidth]{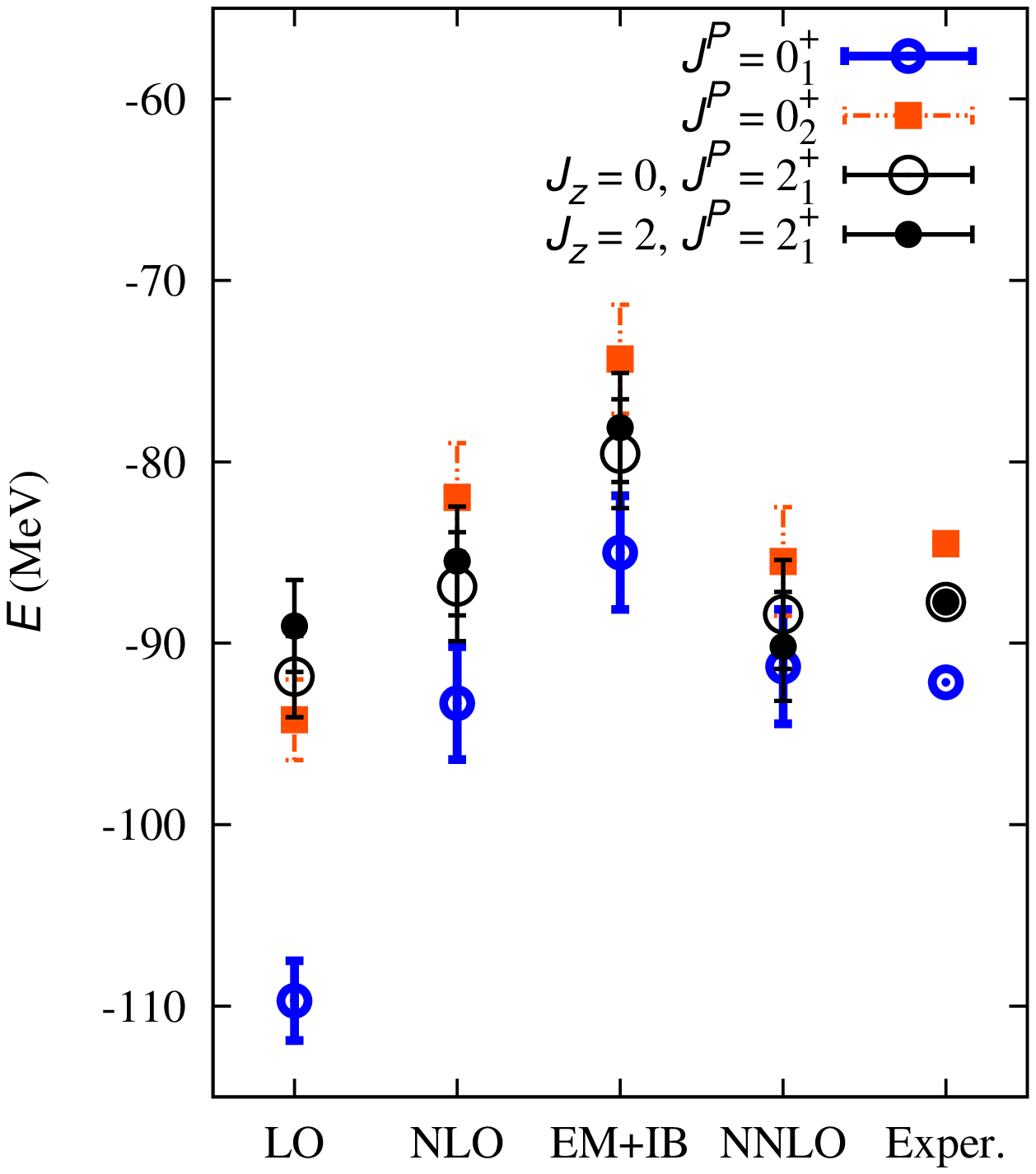}
\caption{\label{fig:hoyle}
Left: Extraction of the excited states of $^{12}$C from the time dependence of
the projection amplitude at LO. The slope of the logarithm of
$Z(t)/Z_{O^+_1}(t)$ at large $t$ determines the energy relative to the ground state.
Right: Summary of the results for the  $^{12}$C spectrum and comparison with
experiment. For each order in chiral EFT labeled on the abscissa,  results are
shown for the ground state (blue circles), the Hoyle state (red squares) and
the $J_z =0$  (open black circles) and  $J_z =2$  (filled black circles)
projections of the lowest-lying spin-2 state.   
}
\vspace{-0.2cm}
\end{center}
\end{figure}
In Fig.~\ref{fig:hoyle} (left panel) the lattice results used to extract the excited state 
energies at leading order are shown. For each excited
state we plot the logarithm of the ratio of the projection amplitudes,
$Z(t)/Z_{0_{1}^{+}}(t)$, at leading order.  $Z_{0_{1}^{+}}(t)$ is the ground
state projection amplitude, and the slope of the logarithmic function at large
$t$ gives the energy difference between the ground state and the excited
state. Note that within the interval of Euclidean time shown, the results 
for the excited states are very clean and thus allow for an unambiguous and
precise extraction of the corresponding energies.
As seen in Tab.~\ref{excited states} and summarized in
Fig.~\ref{fig:hoyle} (right panel), the NNLO results for the Hoyle state
and spin-2 state are in agreement with the experimental values.  While the
ground state and spin-2 state have been calculated in other studies, these 
results are the first \textit{ab initio} calculations
of the Hoyle state with an energy close to the phenomenologically important
$^{8}$Be-alpha threshold.  Experimentally the $^{8}$Be-alpha threshold is at
$-84.80$~MeV, and the lattice determination at NNLO gives $-86(2)$~MeV. We
also note the energy level crossing involving the Hoyle state and the spin-2
state.  The Hoyle state is lower in energy at LO but higher at NLO. One of
the main characteristics of the NLO interactions is to increase the repulsion
between nucleons at short distances.  This has the effect of decreasing the
binding strength of the spinless states relative to higher-spin states.  We
note the $17$~MeV reduction in the ground state binding energy and $12$~MeV
reduction for the Hoyle state while less than half as much binding
correction for the spin-2 state.  This degree of freedom in the energy
spectrum suggests that at least some fine-tuning of parameters is needed to
set the Hoyle state energy near the $^{8}$Be-$^4$He threshold.  It would be
very interesting to understand which fundamental parameters in nature control
this fine-tuning.  At the most fundamental level there are only a few such
parameters, one of the most interesting being the masses of the up and down
quarks. Some investigations have already been performed to unravel the quark
mass dependence of the deuteron binding energy and of the S-wave
nucleon-nucleon scattering lengths \cite{Beane:2002vs,Epelbaum:2002gb}.
The impact on the primordial abundances of light elements created by a 
variation of the quark masses at the time of Big Bang nucleosynthesis was
also studied in Ref.~\cite{Bedaque:2010hr}. Along similar lines, we are 
presently studying the quark mass dependence of the $^{12}$C spectrum.

%%%%%%%%%%%%%%%%%%%%%%%%%%%%%%%%%%%%%%%%%%%%%%%%%%%%%%%%%%%%%%%%%%%%%%

\section{A new look at clustering in nuclear physics}

The cluster model of nuclei, that was initiated by Hastad, Teller and others
many decades ago \cite{teller}, naturally explains the maxima in the
binding energy per particle observed for nuclei with even and equal 
numbers of protons and neutrons. All these nuclei starting with $^8$Be can
be considered as composed of alpha-particles. For $^8$Be, the dumbbell-like
structure emerging from two close-by alphas could be even demonstrated in the
ab initio calculation of Ref.~\cite{Wiringa:2000gb}. Alpha-clustering also 
plays an important role in the Hoyle-state, higher excited states in $^{12}$C
and the analogous states in  $^{16}$O, as discussed by Martin Freer at this
conference \cite{Freer}. In this classical picture, such states are composed
of $\alpha$-particles in various geometrical configurations, e.g. the Hoyle
state is visualized as three loosely bound alphas in a shallow potential,
that leads to weak binding and easy break-up corresponding to $\alpha$-radioactivity.
In such a picture, the Hoyle state has a much larger charge radius than the
ground state, although in more realistic models including configuration
mixing allowing for shell-model components one finds some reduction in 
size~\cite{Chernykh:2007zz}.

The framework discussed here offers a fresh - and may be very different - look
at the phenomenon of clustering. As shown in Fig.~\ref{fig:configs},
configurations with up to four nucleons (two protons and two neutrons with
spin up and down, respectively) on one lattice site are possible. 
Thus, $\alpha$-type
of clustering is inherent in this approach. As alreasdy mentioned
above, if one uses leading order
interactions with no smearing, one encounters a cluster instability, such that
e.g. the ground state of $^4$He is severely overbound and consists almost
entirely of the quantum state with all four nucleons occupying the same
lattice site. This can be understood as the result of two
contributions. First, the LO four-fermion interactions are
momentum-independent and thus too strong at high momenta. Second, there is
a combinatorial enhancement of the contact interactions when more than two
nucleons occupy the same lattice site (for a more detailed discussion, see
\cite{Borasoy:2006qn}). This effect is known from studies
of two-dimensional large-N droplets with zero-range attraction and also
from systems of higher-spin fermions in optical traps and lattices
\cite{Lee:2005nm,Wu:2003zz,Wu:2005zz}. Within chiral EFT, this problem is
cured by either going to higher orders and including the corresponding higher 
derivative operators or, even better, to perform a Gaussian smearing of the
contact interactions with its size tuned to the nucleon-nucleon effective ranges. This
improved LO action that acts differently on the S- and P-waves is then used
in the nonperturbative part of the calculation while all other higher order
operators are treated in perturbation theory. Thus, one effectively has given
the LO contact interactions some extension. Furthermore, as a consequence of
dealing with this phenomenon, we have essentially given a new meaning to 
clustering in nuclei - one does not have to think in terms of classical
$\alpha$-particles but rather in terms of smeared out states with four
nucleons on one lattice site. One might contemplate how this new picture in
the continuum limit $a\to 0$ might recover the old one - such a question can
be addressed in the formulation of lattice chiral EFT of Montvay and Urbach
\cite{MU}, who propose to introduce a momentum cut-off independent of taking
the continuum limit. Clearly, lots of work is required to solidify the arguments
leading to this novel look at clustering, but I am quite confident that the
successful calculation of the $^{12}$C spectrum has already offered us a first hint,
as explained here.
 
%%%%%%%%%%%%%%%%%%%%%%%%%%%%%%%%%%%%%%%%%%%%%%%%%%%%%%%%%%%%%%%%%%%%%%%%%%%%%%%%%
\section{Summary and outlook}

Nuclear lattice simulations are a tool to investigate the structure
of atomic nuclei based on the forces derived from the chiral Lagrangian
of QCD. Besides the the $^{12}$C nucleus and the Hoyle state
discussed here, our collaboration has also
investigated dilute neutron matter in the unitary limit \cite{Epelbaum:2008vj}
and is presently addressing the issue of possible P-wave pairing in neutron
matter. If existing for physical strength of the P-wave interactions,
it would certainly have consequences for the nuclear equation of state and the cooling
rate of neutron stars. Nuclear reaction dynamics can also be addressed
within this framework, however, more conceptual developments are necessary.
A first step in this direction was recently performed. In
Ref.~\cite{Bour:2011ef} it was shown that bound states moving in a finite
periodic volume have an energy correction which is topological in origin and
universal in character. The topological volume corrections contain information
about the number and mass of the constituents of the bound states. These phase
corrections should be considered when determining scattering phase shifts for
composite objects at finite volume (this work has been extended to quantum
field theory in a finite volume in Ref.~\cite{Davoudi:2011md}). Such 
investigations pave the way for
a model-independent calculation of the unsatisfactorily determined reaction
$^{12}C(\alpha,\gamma)^{16}O$ at stellar energies. This process plays an
important role in the element synthesis in stars and in the dynamics
underlying supernova explosions. It is often called the
``holy grail'' of nuclear astrophysics. The framework presented here holds the
promise to solve this outstanding problem and thus might contribute significantly
to our understanding of the generation of the elements needed for life on Earth.

\section*{Acknowledgements}
I thank the organizers for giving me the opportunity
 to present these thoughts at this
wonderful conference. I am grateful to all my
collaborators, who have contributed to my understanding of the
issues discussed here. Partial financial support by the Helmholtz 
Association through funds provided to the virtual institute 
``Spin and strong QCD'' (VH-VI-231), by the European Community-Research 
Infrastructure Integrating Activity ``Study of Strongly Interacting Matter''
(acronym HadronPhysics2, Grant Agreement n.~227431) under the Seventh
Framework Programme of the EU,  by DFG (SFB/TR 16, ``Subnuclear 
Structure of Matter'') and by  BMBF (grant 06BN9006) is gratefully acknowledged.

%\pagebreak

\end{document}